\documentclass[lettersize,journal]{IEEEtran}

\usepackage{amsmath,amsfonts,amssymb}
\usepackage{mathtools}
\usepackage{amsthm}
\usepackage{alltt, xspace, times, epsfig}
\usepackage{gensymb,xfrac,array,booktabs,bm}
\usepackage[linesnumbered,lined,ruled]{algorithm2e}
\usepackage{adjustbox}
\usepackage{booktabs}
\usepackage{soul}  %

\usepackage{mathtools}%
\usepackage{amsmath}

\usepackage[utf8]{inputenc}
\usepackage[english]{babel}

\usepackage{listings}

\usepackage{multirow, multicol, booktabs, tabulary, tabu, longtable, array, varwidth}
\usepackage{placeins, lipsum}
\usepackage[flushleft]{threeparttable}

\usepackage{soul}

\usepackage{tabularx}

\usepackage{amssymb}

\usepackage[labelfont=bf]{caption}
\usepackage{cite}
\usepackage[hyphens]{url}
\usepackage{hyperref}
\usepackage[dvipsnames, table]{xcolor}
\hypersetup{
    linkcolor  = blue!85!black,
    citecolor  = blue!85!black,
    urlcolor   = blue!85!black,
    colorlinks = true,
    breaklinks = true
}

\usepackage{enumerate}

\usepackage[inline]{enumitem}

\usepackage{amsmath, amsfonts, amssymb, amsthm, nicefrac}
\usepackage{mathptmx}  %
\usepackage[mathcal]{eucal}

\theoremstyle{definition}

\theoremstyle{comment}

\usepackage[]{cleveref} %
\crefname{appsec}{Appendix}{Appendices}
\crefformat{section}{\S#2#1#3}
\crefformat{subsection}{\S#2#1#3}
\crefformat{subsubsection}{\S#2#1#3}
\crefformat{equation}{(#2#1#3)}
\Crefformat{equation}{Equation~(#2#1#3)}
\crefrangeformat{equation}{(#3#1#4--#5#2#6)}
\Crefrangeformat{equation}{(#3#1#4--#5#2#6)}
\crefmultiformat{equation}{(#2#1#3)}{ and~(#2#1#3)}{, (#2#1#3)}{ and~(#2#1#3)}
\crefformat{figure}{Fig.~#2#1#3}
\Crefformat{figure}{Fig.~#2#1#3}
\crefmultiformat{figure}{Figs.~#2#1#3}{ and~#2#1#3}{, #2#1#3}{ and~#2#1#3}
\crefformat{algorithm}{Algorithm~#2#1#3}
\Crefformat{algorithm}{Algorithm~#2#1#3}
\crefmultiformat{algorithm}{Algorithm~#2#1#3}{ and~#2#1#3}{, #2#1#3}{ and~#2#1#3}
\crefformat{table}{Table~#2#1#3}
\Crefformat{table}{Table~#2#1#3}
\crefrangeformat{table}{Tables~#3#1#4--#5#2#6}
\Crefrangeformat{table}{Tables~#3#1#4--#5#2#6}
\crefmultiformat{table}{Tables~#2#1#3}{ and~#2#1#3}{, #2#1#3}{ and~#2#1#3}

\usepackage{textcomp}
\usepackage[shortcuts,acronym]{glossaries}
\usepackage{soul, footnote, xargs}
\usepackage[colorinlistoftodos,prependcaption,textsize=tiny]{todonotes}
\usepackage{pifont}
\usepackage{balance}

\newcommand\thefont{\expandafter\string\the\font}

\usepackage{tcolorbox}
\newcounter{remark}

\usepackage{graphicx}
\usepackage[tight,footnotesize]{subfigure}
\usepackage{tikz, epstopdf, stfloats, bbding, capt-of}
\usepackage{algorithmic}
\usepackage[ruled, linesnumbered]{algorithm2e}

\SetCommentSty{mycommfont}
\graphicspath{{figs/}}
\renewcommand{\footnotesize}{\scriptsize}

\usepackage{wasysym}

\newcommand{\sysname}{\textsc{AttackTagger} }

\definecolor{colortx}{HTML}{FABD2F}

\definecolor{textgray}{gray}{0.4}
\usepackage[%
    font={small},
    labelfont=bf,
]{caption}

\usepackage{xstring}

\newcommand{\para}[1]{
\vspace{2px}
\noindent{\bf \IfEndWith{#1}{.}{#1}{#1.}}
}

\newcommand{\revise}[1]{{\color{black} #1}}

\usepackage{breqn}

\usepackage{amsmath}

\begin{document}

\title{Security Testbed for Preempting \\ Attacks against Supercomputing Infrastructure} 

\author{\IEEEauthorblockN{Phuong Cao$^{1,2}$, Zbigniew Kalbarczyk$^1$, Ravishankar K. Iyer$^1$\\}
\IEEEauthorblockA{
\small{$^1$University of Illinois at Urbana-Champaign,
$^2$National Center for Supercomputing Applications.}}
}

\IEEEtitleabstractindextext{%

\begin{abstract}
Securing HPC has a unique threat model. Untrusted, malicious code exploiting the concentrated computing power may exert an outsized impact on the shared, open-networked environment in HPC, unlike well-isolated VM tenants in public clouds. 
Therefore, preempting attacks targeting supercomputing systems before damage remains the top security priority. The main challenge is that noisy attack attempts and unreliable alerts often mask \emph{real attacks}, causing permanent damages such as system integrity violations and data breaches. This paper describes a security testbed embedded in live traffic of a supercomputer at the National Center for Supercomputing Applications (NCSA). The objective is to demonstrate attack \textit{preemption}, i.e., stopping system compromise and data breaches at petascale supercomputers. Deployment of our testbed at NCSA enables the following key contributions: 

1) Insights from characterizing unique \textit{attack patterns} found in real security logs of more than 200 security incidents curated in the past two decades at NCSA. 

2) Deployment of an attack visualization tool to illustrate the challenges of identifying real attacks in HPC environments and to support security operators in interactive attack analyses.

3) Demonstrate the utility of the testbed by running novel models, such as Factor-Graph-based models, to preempt a real-world ransomware family.

\end{abstract}

}

\maketitle

\IEEEdisplaynontitleabstractindextext

\IEEEpeerreviewmaketitle

\maketitle
\section{Introduction} 
\begin{figure*}[t]
    \centering
    \includegraphics[width=\textwidth]{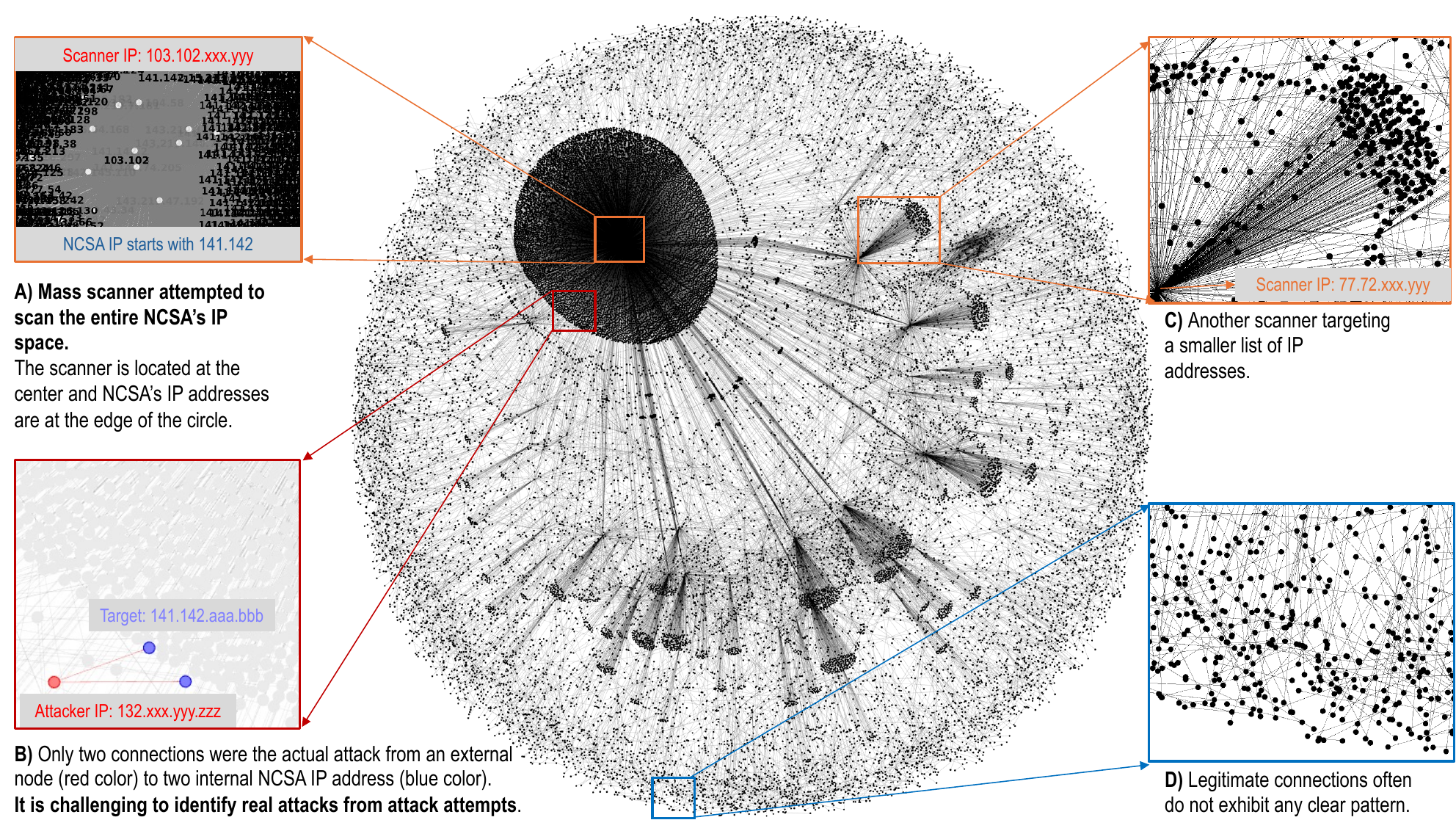}
    \caption{The challenge of finding real attacks hidden in a dense graph of noisy attack attempts intermingle with legitimate connections is illustrated in this graph. \textbf{A}) Most connections are generated by a mass scanner at the circle's center, attempting to scan open network ports across the entire NCSA's \texttt{/16} IP address space (65,536 hosts). \textbf{B}) Real attacks are difficult to find due to the large scale of data, noisy attack attempts from other scanners (\textbf{C}), and legitimate connections (\textbf{D}), which often do not exhibit any clear pattern. \textbf{Data source}: NCSA's black hole router recorded 26.85 million scans on 2024/08/01 from 00:00 to 01:00. We sampled 10,000 most frequent scans from a mass scanner to include in part \textbf{A} in addition to legitimate network connections recorded by Zeek and a real attack. \textbf{Graph drawing method}: The graph contains 29,075 nodes and 27,336 edges and has been rendered using Gephi~\cite{hu2005efficient}. Annotation of attacker nodes has been done manually by cross-examining the ground truth of the attacker's IP addresses provided by the Factor-Graph-based attack detector~\cite{cao2015preemptive,cao2019preempting}.
    }
    \label{fig:network-visualization}
\end{figure*}

Stopping attacks before they cause irreversible damage remains the top security priority. Securing HPC has a unique threat model. Untrusted, malicious code exploiting the concentrated computing power may exert an outsized impact on the shared, open-networked environment in HPC, unlike well-isolated VM tenants in public clouds. 
In high-performance computing (HPC) and supercomputing environments~\cite{peisert2017security}, the main challenge lies in the vast number of intrusion attempts compared to the few successful attacks that lead to system breaches and data leaks. For example, the difficulty of finding real attacks hidden in a dense graph of noisy attack attempts intermingling with legitimate connections is illustrated in a graph visualization in ~\cref{fig:network-visualization}, in which the high volume of data caused by mass scanners overwhelm system operators with alerts, making it difficult to identify real attacks~\cite{sharma2010analysis,pecchia2011identifying}.

This paper addresses the problem of \textit{preempting} attacks, i.e., stopping misuse of HPC resources or data breaches, demonstrated at the National Center for Supercomputing Applications, which hosts a petascale supercomputer with thousands of active users and a comprehensive gamut set of workloads. We present 1) data-driven insights of a longitudinal analysis of a dataset of more than 200 past security incidents at NCSA, curated in 2000-2024, and 2) a testbed architecture (\sysname) and a case study of real ransomware detected in the testbed.

\textbf{Data-driven insights from real-world security incidents}.  Using domain knowledge of security operators on a longitudinal dataset of more than 200 security incidents, we extracted key alert sequences in successful attacks and found the following insights. \textit{The effective range of a detection model is alert sequences with lengths from two to four alerts}. Such attacks have a high degree
of alert similarity. For example, more than 95\% of attacks have up to 33\% similar alerts, which correspond to common attack vectors for establishing a foothold in the
target network before executing exploits to exfiltrate secrets. Similar alert sequences are repeatedly found in old and recent security incidents. This observation indicates
that a detection model built upon matching indicative alert
sequences will have a high chance of catching attacks because
present-day attacks are similar to past attacks. One example of a repeated alert sequence is the pattern: 1) download a source file over an unsecured HTTP connection, 2) compile it as a kernel module, and 3) erase the forensic trace of the attack. This pattern, first observed in 2002, continues to appear in attacks as of 2024 and was found in 60.08\% (137 out of more than 200) of past security incidents in our study, which analyzed 25 million alerts collected in Zeek notice logs over 24 years at NCSA.

Another critical insight is that \textit{critical alerts} such as \textit{privilege escalation} or the \textit{presence of personally identifiable information in an outgoing HTTP request} are strong indicators of data breaches. However, these alerts often appear late in the attack timeline after irreversible damage.  Furthermore, individual alerts may not signal all attacks, as an attacker's behavior is only partially observable. Beyond individual alerts, we found that recurring alert sequences often share common patterns at the onset of an attack, primarily aimed at gaining system access. 

\textbf{Live deployment of the testbed in production network.} The testbed consists of Virtual Machines (VMs) sharing 144 CPU cores, a 400Gbps high-speed network link shared with NCSA's interconnected Terabit-scale backbone optical links, and a dedicated class B IPv4 range that allows $2^{16}$ or 65,536 IPv4 host addresses. This testbed is a successor to our previously deployed Secure Shell (SSH) honeypot at NCSA~\cite{cao2019caudit}. We isolate application VMs in the testbed from NCSA's production system to prevent any accidental impact from injected attacks on scientific experiments. Finally, we continuously \revise{and expose our system to the public Internet, which attracts both naive attacks attempts to exploit remote code execution vulnerabilities and sophisticated threats such as ransomware attacks, to validate different models~\cite{cao2015preemptive,cao2019preempting}}.

\textbf{Main results.} The testbed allows testing new models, such as experimental Factor-Graph-based detectors~\cite{cao2019preempting} on mirrored alerts of all production network traffic at NCSA. As a case study, the deployment of the Factor Graph model on the testbed detected ransomware before humans noticed and notified the security team of a family of its variants from infecting NCSA's production network. 

Our successful detection of the ransomware family results in new security alerts, e.g., lateral movement of the ransomware, being improved and incorporated into Zeek policies in the production network at NCSA. These new alerts are the basis for refining detection models in adapting to future attacks.

\textbf{Contributions:} 

\begin{itemize}
    \item A unique longitudinal measurement study that shows alert patterns of real attacks.
    \item A testbed to evaluate the new attack detector model based on factor graphs to infer hidden attack states and stop attacks before the damage.
    \item Real-world demonstration of real network traffic deployed on a network segment of NCSA's, resulting in the preemption of a family of ransomware attacks.
\end{itemize}

\label{sec:dataset}
\section{Data Set, Graph Visualization, and Insights}
This section describes the dataset of real-world attacks used in this paper, the methodology of visualizing the graph of attacks, key insights gained, and the threat model.

\subsection{Dataset}
Our longitudinal dataset consists of more than 200 security incidents targeting NCSA over 24 years (2002-2024) with a size of approximately 30TB ~\cref{tab:data-overview}. This dataset captures a wide spectrum of attacks, from simple SQL Injections to sophisticated Secure Shell (SSH) keyloggers, ransomware and their variants, and offers an opportunity to generalize them to unseen attacks. The security team at NCSA has forensically examined attack traces in each incident, which include:
\begin{itemize}
    \item human-written incident reports that indicate ground truth: the users and the machines involved in the 
    incident,
    \item raw logs of both legitimate user activities and attack activities, i.e., network flows
    (generated by a cluster of \texttt{Zeek} network security monitors~\cite{paxson1999bro}),
    system logs (generated by \texttt{rsyslog},\texttt{osquery,} and \texttt{ossec}~\cite{bray2008ossec}), and
    \item audit logs of system calls (generated by \texttt{auditd}).
\end{itemize}

The ground truth and raw log data are further annotated and filtered to build and evaluate detection models. 

\textbf{Data pre-processing.} Our dataset has a high volume of alerts, averaging 94K daily, as shown in \cref{fig:alerts-per-month}. To prepare the logs for our model, each log message is assigned a symbolic name indicating the attacker's intention. Specific information (e.g., personal information~\cite{talbi2018towards,chen2017evaluating} or filename ) is sanitized while the log timestamp is kept. For example, the raw log \textit{23:15:22 [internal-host] wget 64.215.xxx.yyy/abs.c  (200 "OK" [7036]} shows a \textit{wget} command downloads a \textit{C} source code file, is represented by a symbol \texttt{alert\_download\_sensitive} and metadata \textit{host: internal-host, source-ip:64.215.xxx.yyy}. This is one of the frequent alert patterns that we will further describe in the key insight section below. Finally, each log message is annotated with metadata indicating the log's origin, such as source IP address or hostname. 

\textbf{Assigning labels to alerts (ground truth).} To effectively annotate attack attempts, true attacks, and legitimate uses, \revise{from the original 25 million alerts~\cref{tab:data-overview}, we filter repeated alerts of periodic scans from the public Internet to reduce the size of our dataset to 191K alerts directly related to successful attacks}.  A majority of alerts (99.7\%) have been automatically annotated with
corresponding attack states. These alerts are benign (e.g., \texttt{login}) or malicious (e.g., installation of a binary file in an existing malware database). Only a small fraction
(0.3\%) of alerts (i.e., ones that appear in both attack and legitimate activities) cannot be
annotated automatically. We consulted with several security experts to annotate the remaining
alerts. While we assume that the annotations by security experts are correct, i.e., attack alerts
are labeled as malicious, we can reuse a body of work in ML that addresses annotation
accuracy~\cite{karger2011iterative,ratner2017snorkel}. 

Once the dataset is annotated, we illustrate the complexity of data at a supercomputing scale through graph visualization and perform analyses to identify key characteristics of alerts appearing in past attacks.

\begin{figure}[!b]
\stepcounter{remark}
\begin{tcolorbox}[width=\linewidth, colback=blue!5!white,colframe=blue!75!black,title=Remark \arabic{remark}: \textbf{Our unique longitudinal dataset provides key insights to characterize HPC attacks and guide the design of detection models}.]
{
\begin{itemize}
    \item Dataset (2000-present) spans a wide attack spectrum, e.g., SSH keyloggers and ransomware.
    \item High fidelity visualization ($\approx$ 27M connections) depicting the challenge of identifying attacks.
    \item Insight on preemption models being effective for attacks with two to four alerts.
\end{itemize}
}
\end{tcolorbox}
\end{figure}

\subsection{Graph visualization of mass scanners and attackers}

The challenge of finding real attacks hidden in a dense graph of noisy attack attempts intermingle with legitimate connections is illustrated in~\cref{fig:network-visualization}. The key parts of the figure are: \textbf{A}) Most connections are generated by a mass scanner at the circle's center, attempting to scan open network ports across the entire NCSA's \texttt{/16} IP address space (65,536 hosts). \textbf{B}) Real attacks are difficult to find due to the large scale of data, noisy attack attempts from other scanners (\textbf{C}), and legitimate connections (\textbf{D}), which often do not exhibit any clear pattern. The figure is generated as follows.

\textbf{Data source}: NCSA's black hole router recorded 26.85 million scans on 2024/08/01 from 00:00 to 01:00. We sampled 10,000 most frequent scans from a mass scanner to include in part \textbf{A} in addition to legitimate network connections recorded by Zeek and a real attack. The sample size of 10,000 was experimentally chosen because increasing the size of our sample would not increase the fidelity of the figure since the scans were already visualized as a dense mass. An example of connection data is shown below in the Graphviz file format ~\cite{ellson2002graphviz}. Each line shows a source IP address connecting to a destination IP address. Only the first part of an IP address is shown to preserve privacy, as listed in the verbatim log below.

\begin{lstlisting}
digraph { 
    194.28. -> 143.219.
    71.201. -> 143.219.
    72.21. -> 141.142.
    64.39. -> 141.142.
    103.102. -> 141.142.
    103.102. -> 141.142.
    ...
    103.102. -> 141.142.
    103.102. -> 141.142.
    216.158. -> 141.142.
    91.247. -> 143.219.
}
\end{lstlisting}

\textbf{Graph drawing method}: The graph contains 29,075 nodes and 27,336 edges and has been rendered using Gephi~\cite{hu2005efficient}. Annotation of attacker nodes has been done manually by cross-examining the ground truth of the attacker's IP addresses provided by the Factor-Graph-based attack detector. For example, the mass scanner's IP address \texttt{103.102} is provided by NCSA's Black Hole Router, indicating a cloud provider from Indonesia scanning internal NCSA nodes \texttt{141.142}.

In sum, this graph visualization provides a unique perspective on attack attempts in a real HPC environment at NCSA.

\begin{table}[!b]
\centering
\begin{tabular}{@{}|l|l|@{}}
\toprule
\textbf{Data}               & \textbf{Size} \\ \midrule
\begin{tabular}[c]{@{}l@{}}Total alerts related to\\ successful attacks\end{tabular} & 25 M \\ \midrule
Alerts after being filtered & 191 K\\ \midrule
Successful attacks          & more than 200 incidents \\ \midrule
Data size                   & 30 TB         \\ \midrule
Time period                 & 2000-2024     \\ \bottomrule
\end{tabular}
\caption{Overview of our security incidents dataset (2000-2024)}
\label{tab:data-overview}
\end{table}

\subsection{Key Insights}
Using our dataset, ground truth, and interactive visualization graph above, we identified key insights that characterize past attacks.

\begin{figure*}[!ht]
    \centering
    \begin{minipage}{.45\textwidth}
    \centering
    \includegraphics[width=\columnwidth]{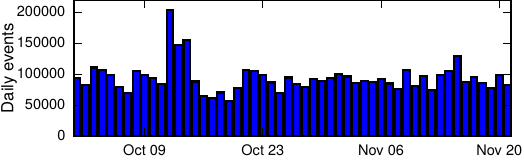}
    \caption{NCSA's monitors observe an average of 94,238 alerts per day (standard deviation = 23,547) in a sample month.}
    \label{fig:alerts-per-month}
    \end{minipage}%
\hfill
    \begin{minipage}{.45\textwidth}
    \centering
    \includegraphics[width=\columnwidth]{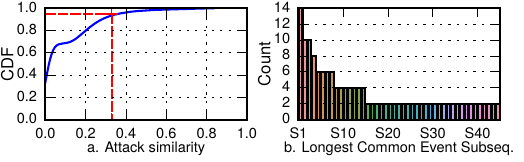}
	\caption{(a) The fractions of similar alerts between pairs of attacks in our dataset. (b) The count of LCS in our dataset.}
    \label{fig:attack_similarities}
    \end{minipage}%
\hfill

\end{figure*}

\textbf{Insight 1: Attacks in our dataset have a high degree of alert similarity}. ~\cref{fig:attack_similarities}-a shows the pair-wise Jaccard similarity of common alert sequences~\cite{nistlcs} between attacks plotted in a cumulative distribution function. For example, the red line shows that more than 95\% of attacks have up to 33\% of similar alerts indicative of attacks. These alerts correspond to common attack vectors for establishing a foothold
in the target network before executing exploits to exfiltrate secrets. We found that similar alert sequences are repeatedly found in old and recent security incidents. This observation indicates that a detection model built upon matching indicative alert sequences will have a high chance of catching attacks because present-day attacks are similar to past attacks.

\textbf{Insight 2: The effective range of a detection model is alert sequences with lengths from two to four alerts.} We have identified common alert sequences (name from S1 to S43, which we will release in the Appendix upon publication of the paper) occurring in past attacks and plotted the frequency at which we saw them in ~\cref{fig:attack_similarities}-b. These sequences have a length from two up to fourteen alerts. For example, the histogram indicates that the most frequent attack pattern (S1) has been seen 14 times across more than 200 past security incidents.

We analyzed the length of alert sequences in such common alert sequences. Attacks that exhibit only one alert are related to sudden attacks: such as a successful zero-day remote code execution that occurred instantly and thus cannot be preempted by a model. On the other hand, attacks with alert sequences larger than or equal to five alerts mean the attack has matured and caused damage, lessening the utility of any detection model. As such, an attack preemption model must work with sequences of two to five alerts to detect the attack. This is the case for attacks that are in progress.

\textbf{Insight 3: The attack sophistication can be indicated by the timing of recurrent alerts.} Attacks the wild~\cite{aptnotes} often start with a set of
\emph{repetitive} but \emph{inconclusive} alerts to identify vulnerable computing resources (e.g.,
scans for vulnerable Apache Struts portals~\cite{equifax}). Such repeated alerts constitute most daily alerts: 80K out of 94K in \cref{fig:alerts-per-month} are repeated port and
vulnerability scans daily. These alerts are not
indicators of malicious activity but can be used to signal potentially malicious alerts that need further monitoring. On the other hand, we observe that once an attacker identified a target, they would manually carry out the attack. Thus, the time between alerts in this stage exhibits significant variability, influenced by system configuration, the specific exploit used, and
the attacker’s approach to analyzing results. 

\textbf{Insight 4. The quantification of alert criticality must be considered during detection.}
\emph{Critical alerts} can be used to detect successful attacks; however, they cannot be used to preempt
attacks because their occurrences indicate that the system integrity has already been
compromised and that data have already been exfiltrated~\cite{bonaci2015experimental}. The entire
dataset has 19 such unique critical alerts, which occur 98 times in the more than 200 attacks. \revise{In cases where
critical alerts were recorded, it was too late to preempt the system integrity loss. On the other
hand, if any of the alerts were considered as an indicator of a complete attack, then analysts would have to analyze all of the alerts (e.g., $94K$ daily alerts
observed at NCSA in \cref{fig:alerts-per-month}) which is impractical.}

\begin{figure}[!b]
\stepcounter{remark}
\begin{tcolorbox}[width=\linewidth, colback=blue!5!white,colframe=blue!75!black,title=Remark \arabic{remark}: \textbf{A preemption model must consider the conditional probability of an alert when making a decision}.]
{
The key characteristics of observed alerts are: 
\begin{itemize}
    \item When a critical alert such as unauthorized privilege escalation is observed, it is certain that an attack is underway; however, the damage is irreversible (i.e., data has been exfiltrated), or
    \item Most daily attack attempts and mass brute-force scans will fail. Thus, their false positive rate is high.
    \item An accurate model must incorporate conditional probabilities of an alert being in a successful attack and normal operational conditions to make a decision.
\end{itemize}
}
\end{tcolorbox}
\end{figure}

\section{Threat model and Related Work}
\revise{This paper considers attacks targeting an open-network environment~\cite{sharma2010analysis}, typically employed in supercomputing/HPC centers, in which multiple users log in through Secure Shell and execute jobs on internal hosts (physical/virtual machines)}. Users can bring their own code and execute arbitrary, including malicious, code~\cite{peisert2017security}. The system is assumed to be benign at the onset. However, the system may have unpatched vulnerabilities or exposed services, which are bound to be exploited due to the complexity of patching highly
interconnected system components~\cite{pashchenko2018esem,lauinger2017thou}. Here, we define key terms in this paper:

\subsection{Key alert concepts: successful attacks, attack attempts, and significant alerts.} We summarize key alert concepts, some of which have been characterized in our recent paper~\cite{yang2024true}. We consider successful attacks as intrusions that caused system compromise and/or resulted in data leaks that are described in a \textit{security incident report} that contains snippet logs of attacks. An attack attempt such as brute-force password scans, significant alerts worth paying attention to such as the download of a known malicious backdoor, and most importantly, the \textit{ground truth} containing the IP address and/or identifier of compromised users and machines. 

\subsection{Attacker and defender capabilities}
\textit{Defender capabilities.} We assume that the alerts from network- and kernel-based monitors are
trustworthy~\cite{garcia2011diversity,beham2013intrusion,babay2018network,liu2018towards,king2005enriching,lee2013high,ma2016protracer,xia2012cfimon,kurmus2011attack,hossain2015towards} and accurate in
capturing attack activities. Since our approach's accuracy depends on monitors, we use an extensive
set of well-configured (e.g., NCSA uses a Zeek cluster for network monitoring) and well-protected
monitors (e.g., osquery runs at the kernel level). While an attacker may tamper with one monitor (on one
host) by using the credentials of a local privileged user, it would be challenging to manipulate \emph{all} monitors. 

\textit{Attacker capabilities.}
We consider that an attacker can pretend to be a legitimate user by using weak/stolen
credentials~\cite{mazurek2013measuring,egele2017towards,barronhoney,han2016shadow} or remote command
execution exploits~\cite{CVE-2017-5638} to compromise internal hosts. \sysname treats it as a single attack if (1) an attacker moves laterally (e.g., connects by SSH to multiple machines) using the same user account and (2) multiple (coordinated or independent) attackers launch an attack using the same user account. If (1) an attacker moves laterally using different user accounts, or (2) one or more attackers use different entry points and launch attacks using different user accounts, \sysname treats that as multiple separate attacks.

\subsection{Scope of this paper} 

This paper describes a testbed infrastructure (\sysname) to support the evaluation of preemption models focused on detecting various attacks targeting supercomputing/HPC environments described in~\cite{sharma2010analysis}. 

We note some limitations and the scope of operation for such models as follows. A preemption model cannot preempt an attack if the attacker (e.g., a malicious insider) executes an attack  
\begin{enumerate*}[label=(\roman*)]
    \item in a single step, immediately causing damage, without being persistent,
    \item with no time evolution involving prior alerts, and
    \item without alerts in common with any of the past network intrusions.
\end{enumerate*}

\section{Testbed architecture and deployment}

\begin{figure*}[ht!] 
    \centering
    \includegraphics[width=\textwidth]{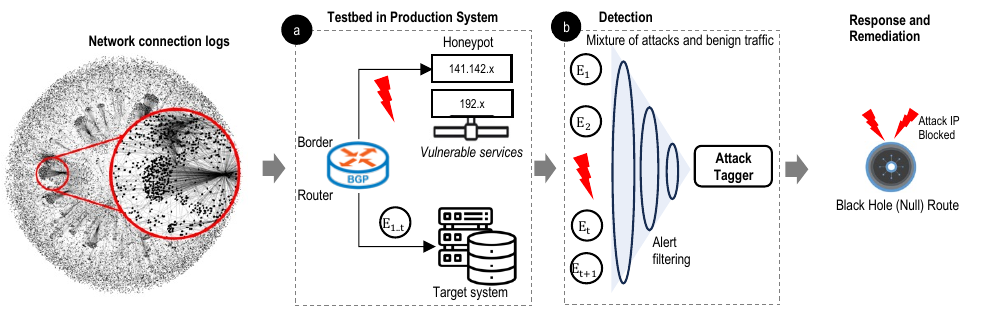}
    \caption{\textbf{Testbed workflow and architecture.}}
    \label{fig:testbed}
\end{figure*}

This section describes the testbed architecture (~\cref{fig:testbed}) that supports embedding a honeypot in a production system to attract attacks, filtering and forwarding alerts into various detection models, e.g., rule-based detector~\cite{cao2015preemptive} or factor-graph-based detector~\cite{cao2019preempting}, and finally, interfacing with a Black Hole router~\cite{cao2019caudit} through automated/programmable Application Programming Interface (API) of the Black Hole Router~\cite{ncsabhrc77:online} for real-time response to mass scanners and remediation of sophisticated/targeted attacks. We describe the key testbed components as follows.

\subsection{Reproducing vulnerable services in a honeypot} 
To attract attackers, we created vulnerable services such as databases that are vulnerable to default passwords or contain remote code execution bugs. The purpose of such vulnerable database hosting infrastructure embedded in production networks is to attract sophisticated attacks such as ransomware (see Fig. \ref{fig:testbed}). Such vulnerable services are hosted on a separate network segment, i.e., a honeypot, that needs a strict isolation policy, i.e., executions are monitored so that malicious traffic does not escape to the internal network or to the Internet. The commands issued by attackers must be closely monitored by the host and network monitors such as Zeek or Osquery.

The key challenge in creating a testbed is to replicate vulnerable software. New Linux distributions, e.g., RedHat or Ubuntu, already patch known vulnerabilities out of the box. A straw man approach of compiling an old, vulnerable package version does not work because of incompatible dependencies. A successful compilation of an old package depends on a specific set of corresponding dependencies unavailable in the latest Linux distributions. As an example, to reproduce the Heartbleed vulnerability, one would need to obtain: i) an old Linux distribution released just before the vulnerability announcement date (April 1, 2014), ii) an old version of the vulnerable \texttt{openssl 1.0.1-f} package, and iii) all dependent packages for building the vulnerable openssl package. 

To address this challenge, we built a \emph{vulnerability reproduction} (VRT) tool~\cite{CSLDepen6:online} to create old Linux containers at any point in the past (2005--present) using the Debian Linux \emph{snapshot} repository\footnote{https://snapshot.debian.org/archive/debian/yyyymmddT000000Z}. The tool is a set of scripts that takes a date, e.g., 20140401, as input. The scripts automatically find a suitable Debian Linux distribution image, e.g., Debian 7 (wheezy), released just before the input date. Then, the scripts configure the image, using Debian's \texttt{debootstrap} utility, to refer to the Debian snapshot repository that contains all the dependencies for that particular date. These scripts work because, since 2005, the Debian project has maintained an archive of such dependencies, i.e., a daily snapshot of dependencies. 

This tool~\cite{CSLDepen6:online} has been in use since 2016 and allows us to reproduce various database attack scenarios, including i) a vulnerable server container with an unpatched kernel, application databases, and their dependencies, and ii) an attack container with corresponding exploits (e.g., ICMP tunneling tools). Thus, our testbed is more extensible than Metasploit, which only provides exploit code and scenarios for memory error vulnerabilities on the host.

\subsection{Attracting attackers}
The attackers are naturally attracted to NCSA's IP address space due to its history of hosting open science research. In addition, we attract attackers by publicly advertising default or user-generated access credentials, e.g., a default "admin" password or a ghost account in a federated identity provider. These “hints” (credentials, database URL, and path) are accidentally published online via various channels such as social media or git. Our database is semi-open, meaning the database is not leaked and is not indexable by search engines, but the credentials are. The use of unique user-generated access credentials (keys) allows us to trace an individual attacker's tactics through the use of user-generated access credentials (keys).

\subsection{Deploying vulnerable services in the live network}
The final deployment state of the honeypot is to host the entry point to the honeypot on a segment of NCSA's network, listening on specific ports, to forward incoming traffic to a private cloud that runs actual instances of the virtual machines/containers. The database, e.g., PostgreSQL, containing the above-generated data, is embedded in the network infrastructure of a large-scale scientific computing network (more than 13,000 computing nodes). We allocated a dedicated \texttt{/24} IP space containing sixteen entry points to such a database. Each entry point is a Virtual Machine that forwards incoming traffic to an isolated container containing the vulnerable or semi-open database. Multiple instances of the database are scaled using Linux containers to cast a wide net, increasing the chance of attracting ransomware.

\begin{figure}[!b]
\stepcounter{remark}
\begin{tcolorbox}[width=\linewidth, colback=blue!5!white,colframe=blue!75!black,title=Remark \arabic{remark}: \textbf{Succesful testbed deployment requires tools supporting reproducible attacks and interactive visualization}.]
{
\begin{itemize}
    \item The testbed architecture offers invaluable insights and a blueprint for implementing comparable services in other HPC environments.
    \item Our tools (reproducing vulnerabilities, black hole router) have been used at NCSA and open-sourced at ~\cite{NCSA45:online}. 
    \item The graph visualization data and attack patterns will be released upon publication of this paper.
\end{itemize}
}
\end{tcolorbox}
\end{figure}

To reduce the risk that an attacker may escape the honeypot, we simultaneously apply the following strategies. First, we provide an immutable virtual machine image to be launched on the testbed. Each instance of the virtual machine is short-lived to reduce the risk of permanent compromise and can be quickly provisioned after successfully collecting malicious traces of attacks. This setup also allows auto-scaling of a network of virtual machine instances, e.g., simulating a distributed federation of databases, allowing us to capture realistic lateral movement attacks. Second, to ensure full control over the attacks, we contained each database in a Linux container and 
further encapsulated vulnerable containers in a QEMU virtual machine with limited capabilities. All containers in our experiments ran in a network sandbox that implemented a Layer-3 private overlay network on a separated Classless Inter-Domain Routing block. In addition, we set up \emph{iptable} rules on container hosts to monitor all new outgoing connections and drop them before their packets were routed to the Internet. 

The testbed is deployed on two dedicated machines with an out-of-band and firewalled network link to maintain in-depth monitoring by Osquery and Zeek.

\section{Case Study: \\ Successful Ransomware Detection in Live Traffic}
The testbed has been deployed with several preemption models ~\cite{cao2015preemptive,cao2019preempting}. The honeypot component of the testbed successfully attracted a real ransomware family attempting lateral movement within the NCSA network. This ransomware family has been detected by our preemption model. We use this case study to illustrate how our testbed provides an early warning to security operators at the National Center for Supercomputing Applications, enabling them to prepare for potential attacks.

The key steps of the ransomware attack attempts are described below.

\subsection{Initial entry of the ransomware}
There have been repeated probing of PostgreSQL database ports in October, and on October 30th, the ransomware entered through an open port \texttt{5432} of a PostgreSQL instance containing the above ML-generated data. As the attacker has privileged access to the database, they perform the following steps.

\begin{itemize}
\item Step 1. Reconnaissance by querying the server version through command \texttt{SHOW server\_version\_num} to find potential vulnerabilities.

\item Step 2. Encoding malicious payload into a PostgreSQL \texttt{largeobject} data type, which can be embedded in the database directly. Here, the binary content of a file is encoded as a hex string \texttt{7F454C46..}. Note that, the prefix of this hex string indicates the magic number of an Executable and Linkable Format (ELF) executable starting with the byte 7F followed by "ELF" (7F 45 4C 46).

\item Step 3. Creating malicious files \texttt{/tmp/kp} on the disk to be executed through the \texttt{io\_export} command.

\end{itemize}
The ransomware then attempted to spread laterally by enumerating known hosts and using stolen SSH keys to deploy the malicious payload automatically. Several malicious files were dropped onto the disk of the PostgreSQL instance, which could have been executed to steal information and further spread the ransomware.

The deployed preemption model~\cite{cao2015preemptive,cao2019caudit} detected the ransomware's communication with its command and control server and immediately notified security operators. This early detection proved crucial, as the actual attack and malicious scripts were observed just twelve days later, confirming the accuracy of our prediction.

\begin{figure}[t] 
    \centering
    \includegraphics[width=0.5\textwidth]{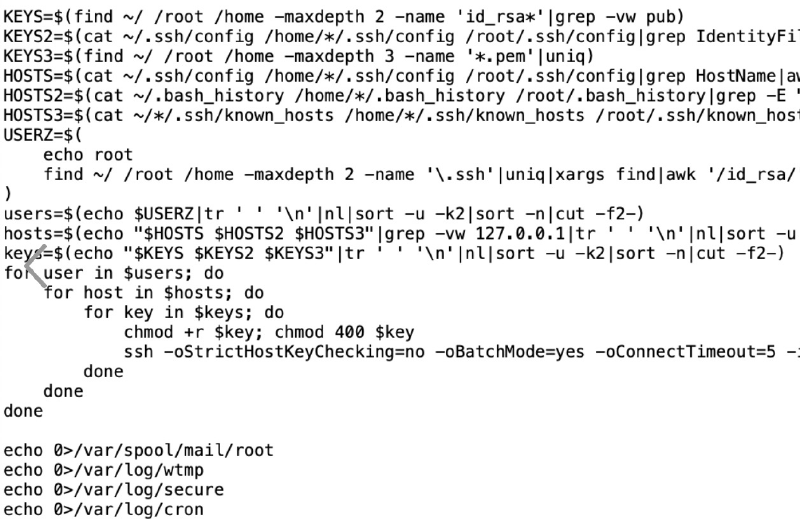}
    \caption{\textbf{Recursive lateral movement of the ransomware by enumerating known hosts in the compromised instance.}}
    \label{fig:lateral}
\end{figure}

\begin{figure}[!hb]
\stepcounter{remark}
\begin{tcolorbox}[width=\linewidth, colback=blue!5!white,colframe=blue!75!black,title=Remark \arabic{remark}: \textbf{A ransomware family has been attracted by our honeypot and preempted by our model~\cite{cao2019preempting}}.]
{
\begin{itemize}
    \item The ransomware attempted to spread its infection laterally using stored Secure Shell (SSH) keys.
    \item Our preemption model notified a security team twelve (12) days early before another similar ransomware targeted NCSA infrastructure.
\end{itemize}
}
\end{tcolorbox}
\end{figure}

\subsection{Lateral movement after installation of malicious files.}
The ransomware then moved recursively by enumerating known hosts in the compromised instance.  Fig. \ref{fig:lateral} shows This malicious payload is crucial for the ransomware to spread exponentially inside the network containing the compromised instance. From the compromised machine, all secret data related to private secure shell (SSH) keys are enumerated and collected using bash scripts. Then, a loop instructs the script to connect to each historical (known) host that the initial compromised machine was authorized to connect. This script uses the above stolen SSH keys in a batch mode to automatically spread the malicious payloads to the target known hosts.

\subsection{Successful detection and notification to security operators.}
Our model detected the ransomware upon its attempt to communicate with its command and control server. Immediately after the observation of the above attack attempt, we notified the security operators at the National Center for Supercomputing Applications so they to be prepared for similar attacks. On Nov 10th, twelve days after our notification, the exact attack and malicious scripts were observed and recorded in a security incident. The initial connection to port \texttt{5432} of PostgreSQL and SSH behaviors was consistent with our motivating ransomware above. This result demonstrates the power of our approach in preempting the attacks.

\begin{lstlisting}[language=bash]
Alerted to the following downloads to this host at 3:44a
hXXp://194.145.xxx.yyy/sys.x86_64 
hXXp://194.145.xxx.yyy/ldr.sh?e7945e_postgres:postgres
We were also alerted an hour later to SSH scanning 
behavior coming from this host. 
...
As best as I can tell, initial malicious connection was 
from 111.200.zzz.ttt connecting to postgres service 
(5432/tcp) on this host.
\end{lstlisting}

In summary, the testbed has successfully supported preemption models for early prediction of a ransomware family that our honeypot has attracted.

\section{Conclusion}\label{conclusion}
Our testbed has been deployed in live traffic of a supercomputer at the National Center for Supercomputing Applications (NCSA), providing a valuable blueprint and early warning of sophisticated attacks such as ransomware before causing a data breach or system misuse. The key contributions are:

1) Insights from mining unique \textit{attack patterns} found in real security logs of more than 200 security incidents curated in the past two decades at NCSA. 

2) Deployment of an attack visualization tool to illustrate the challenges of identifying real attacks in HPC environments and to support security operators in interactive attack analyses.

3) Demonstrate the utility of the testbed by running probabilistic graphical models to detect a real-world ransomware family.

The core tools have been open-sourced on NCSA Github~\cite{NCSA45:online}, and a sample dataset will be released upon publication of this paper.

\section{Acknowledgements}
The authors would like to thank NCSA staff for continuous support in data analysis and testbed deployment; Coordinated Science Lab, Department of Electrical and Computer Engineering, and Siebel School of Computing and Data Science faculty and students for engaging in discussions and initial measurement of attack attempts; NSF for funding part of the main author's research as part of the XSEDE (Extreme Science and Engineering Discovery Environment) program; the TrustedCI leadership team; and other TrustedCI fellows and community members for their valuable feedback and support. 

{
    \balance
    \bibliographystyle{IEEEtran}
    \bibliography{IEEEabrv,references,phuong}
}
\end{document}